\documentclass[a4paper,11pt]{article}
\usepackage{pos}

\DeclareMathOperator{\tr}{tr}
\DeclareMathOperator{\sign}{sign}
\title{QCD thermodynamics with dynamical chiral fermions}
\ShortTitle{QCD thermodynamics}

\author[a,b,c,d]{Z.~Fodor}
\author*[a]{A.Yu.~Kotov}
\author[c,e]{T.G.~Kovacs}
\author[a,b]{K.K.~Szabo}

\affiliation[a]{
Jülich Supercomputing Centre, Forschungszentrum Jülich\\ D-52428 Jülich, Germany}

\affiliation[b]{
Department of Physics, University of Wuppertal\\D-42119 Wuppertal, Germany}

\affiliation[c]{
Institute of Physics and Astronomy, ELTE E\"otv\"os Lor\'and University \\
  P\'azm\'any P\'eter s\'et\'any 1/a,  H-1117 Budapest, Hungary
}

\affiliation[d]{
Physics Department, Pennsylvania State University\\ University Park, PA 16802, USA
}

\affiliation[e]{
Institute for Nuclear Research (ATOMKI), \\ 
H-4026 Debrecen, Bem t\'er 18/c, Hungary
}

\emailAdd{fodor@bodri.elte.hu}
\emailAdd{a.kotov@fz-juelich.de}
\emailAdd{tamas.gyorgy.kovacs@ttk.elte.hu}
\emailAdd{k.szabo@fz-juelich.de}

\abstract{We discuss properties of thermal Quantum Chromodynamics obtained by means of lattice simulations with overlap fermions. This fermion discretisation preserves chiral symmetry at finite lattice spacing. We present details of the formulation and results for the chiral observables. We determine the topological susceptibility from simulations at fixed global topological charge based on the slab method. Using the measured values of the topological susceptibility we sum the chiral observables over all topological sectors. The volume dependence of the chiral susceptibility is in agreement with the crossover nature of the thermal QCD phase transition. Additionally we discuss the spectrum of the overlap Dirac operator and its volume and temperature dependence. Presented results are obtained at the temporal lattice extent $N_t=8$.}

\FullConference{The XVIth Quark Confinement and the Hadron Spectrum Conference (QCHSC24)\\
 19-24 August, 2024\\
 Cairns Convention Centre, Cairns, Queensland, Australia\\}

\begin{document}
\maketitle

\section{Introduction}

QCD thermodynamics has been studied on the lattice quite extensively, for a recent review see, e.g. \cite{Aarts:2023vsf}. One of the most important results is the nature of QCD phase transition at finite temperature and vanishing baryon number, which turns out to be a crossover \cite{Aoki:2006we}. This result was obtained with sometimes debated staggered fermion discretisation, with has only non-single $U(1)$ remnant of the full chiral symmetry and requires questionable rooting procedure.
In this proceeding we discuss our study of QCD thermodynamics around the chiral transition by means of the overlap fermion discretization. Its main advantage is the presence of the chiral symmetry at finite lattice spacing \cite{Neuberger:1997fp}. First results of our study were reported in \cite{Fodor:2024gbe}.

\section{Lattice details}
Below we present the details of the action. We follow the approach, used in \cite{Borsanyi:2016ksw,Borsanyi:2015zva} and previously in \cite{Borsanyi:2012xf}. Note that the generation of configurations is done with fixed global topological charge.

Our lattice setup consists of the following fields:
\begin{itemize}
    \item gluon field with tree-level Symanzik improved lattice gauge action;
    \item $2+1$ flavours of overlap fermions, coupled to 2 steps of HEX smeared \cite{Durr:2007cy} gauge links:
    \begin{equation}
    	aD_{\mathrm{ov}}=\frac{1}{2}(1+\gamma_5\sign(\gamma_5 D_{\mathrm{W}}(-m_W))),
    \end{equation}
    where for the kernel the Wilson-Dirac operator $D_{\mathrm{W}}$ with negative mass $-m_w=-1.3$ is used;
    \item $2$ flavours of Wilson fermions coupled also to the 2-HEX smeared gauge links. This Dirac operator is identical to the one we use as the kernel of the overlap operator;
    \item $2$ additional boson fields with mass $m_B=0.54$ and action:
    \begin{equation}
    	S=\phi^{\dag} \left(D_{\mathrm{w}}(-m_W)+im_B\gamma_5\tau_3\right)\phi.
    \end{equation}
\end{itemize}

The last two fields were introduced in order to suppress the tunnelings between different topological sectors to keep the Monte Carlo simulations in one fixed topological sector $Q=\mathrm{const}$. These fields become irrelevant and decouple in the continuum limit \cite{Fukaya:2006vs}. Since we are doing simulations in fixed topological sectors, we can use the trick described in \cite{Borsanyi:2016ksw} to simulate $1$ overlap fermion flavour, required for the strange quark. We use the LCP, calculated in \cite{Borsanyi:2016ksw} with $N_f=3$ quark flavours, using additional simulations with heavy staggered fermions. In \cite{Fodor:2024gbe} this LCP was also extended to lower $\beta$ values, needed for simulations presented later on.

The inversion of the overlap Dirac operator is performed using the flexible GMRES method \cite{Brannick:2014vda} with the inverted clover Wilson Dirac operator as preconditioner. The inversion of the Wilson Dirac operator is done with the BiCGSTAB method.

We generated $O(500-1000)$ trajectories per aspect ratio $N_s/N_t$ in different topological charge sectors $Q$.

\section{Topological susceptibility from simulations in a fixed topological sector}
\label{sec:slab}
Since our dynamical overlap simulation algorithm cannot tunnel between different topological charge sectors, we cannot directly measure the topological susceptibility. However, even in fixed sector simulations, local fluctuations of the topological charge make it possible to determine the topological susceptibility using the slab method \cite{Bietenholz:2015rsa}. Below we briefly summarize this method.

\begin{figure}
    \centering
    \includegraphics[width=0.5\linewidth]{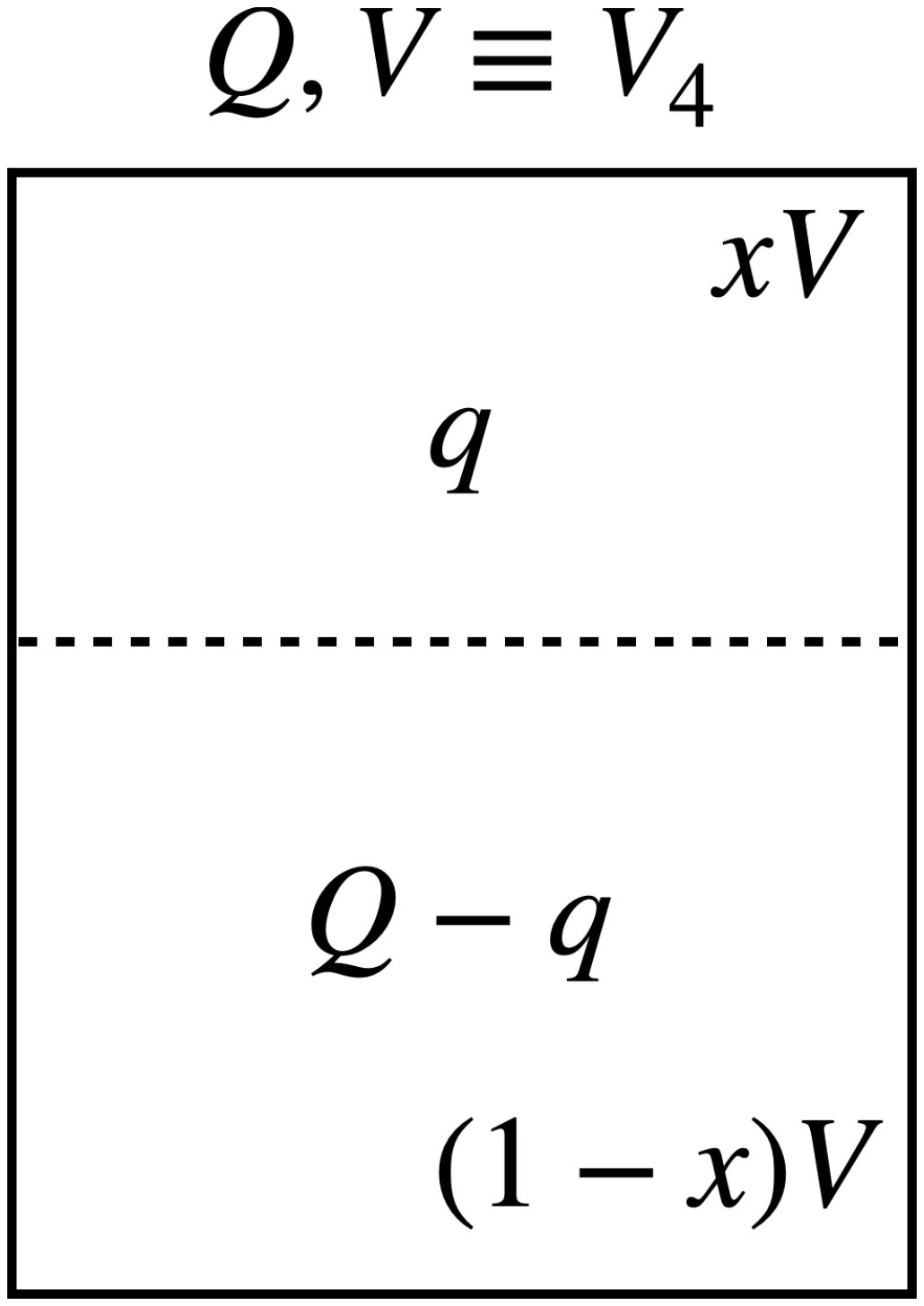}
    \caption{Setup of the slab method. The global topological charge if fixed to $Q$, the topological charge of the subvolume $xV$ is $q$, in the rest part of the volume $(1-x)V$ topological charge is $Q-q$.}
    \label{fig:slab}
\end{figure}

Let us assume that the simulations are done in the (four-dimensional) volume $V$ with fixed value of the topological charge $Q$. If we consider a subvolume $xV$ of the whole volume, then due to local fluctuations of the topological charge, the value of the topological charge $q$ in this subvolume is not fixed. In the rest of the volume $(1-x)V$ the topological charge is then equal to $Q-q$. This situation is visualized in the Fig.~\ref{fig:slab}. Assuming that the volumes are large enough and  the total charge is fixed to be $Q$, the probability of having topological charge $q$ in the subvolume $xV$ can be expressed as
   
\begin{equation}
    p_Q(q)\sim p_1\left(q,xV\right)p_2\left(Q-q,(1-x)V\right)\sim e^{-\frac{q^2}{2\chi xV}}e^{-\frac{(Q-q)^2}{2\chi (1-x)V}}\sim e^{-\frac{q'^2}{2\chi Vx(1-x)}},
\label{eq:slab}
\end{equation}
  
where $q'\equiv q-xQ$, $p_1\left(q,xV\right)$ and $p_2\left(Q-q,(1-x)V\right)$ are the probabilities to find the topological charge $q$ or $Q-q$ in the volume $xV$ or $(1-x)V$ correspondingly. Eq.~(\ref{eq:slab}) implies, that if we plot average $\langle q'^2\rangle$ as a function of $x$, the ratio of the studied subvolume to the total volume, it should follow the parabola behaviour given by $x(1-x)\chi V$. Following \cite{Bietenholz:2015rsa}, we fit a function of the form  $a x(1-x)+b$ to $\langle q'^2\rangle$ in some range around $x\sim0.5$ and extract the topological susceptibility from the value of $a$.

In our lattice simulations we cut the full volume along one spatial direction into two subvolumes. In order to measure the topological charge $q$ in the subvolume, we use the overlap-based definition of the topological charge density, $q(z)=1/2\tr_{\mathrm{c,d}} (\gamma_5 D_{\mathrm{ov}}(z,z))$ and $q=\sum_{z\in xV} q(z)$ integrated over the given subvolume. Here the trace $\tr_{\mathrm{c,d}}$ is taken over color and Dirac indices.  We calculated the trace of the overlap Dirac operator $D_{\mathrm{ov}}(z)$ using the stochastic trace estimator with $O(32-512)$ random sources.

\begin{figure}
    \centering
    \includegraphics[width=0.5\linewidth]{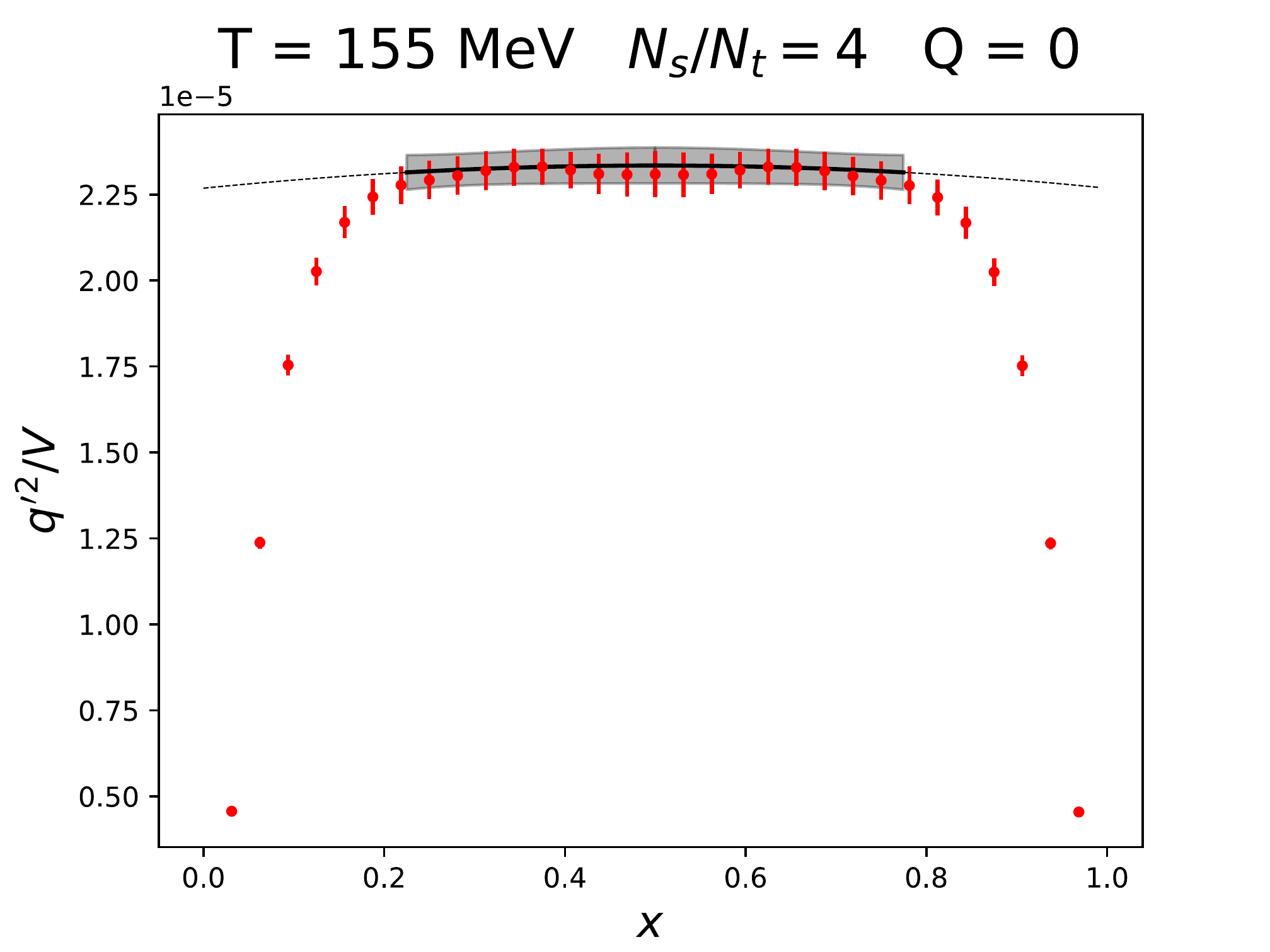}
    \caption{Example of the $\langle q'^2\rangle$ as the function of the subvolume $x$. The temperature is $T=155$~MeV, aspect ratio $N_s/N_t=4$, the global topological charge is $Q=0$.}
    \label{fig:slabcorrelator}
\end{figure}

\begin{figure}
    \centering
    \includegraphics[width=0.7\linewidth]{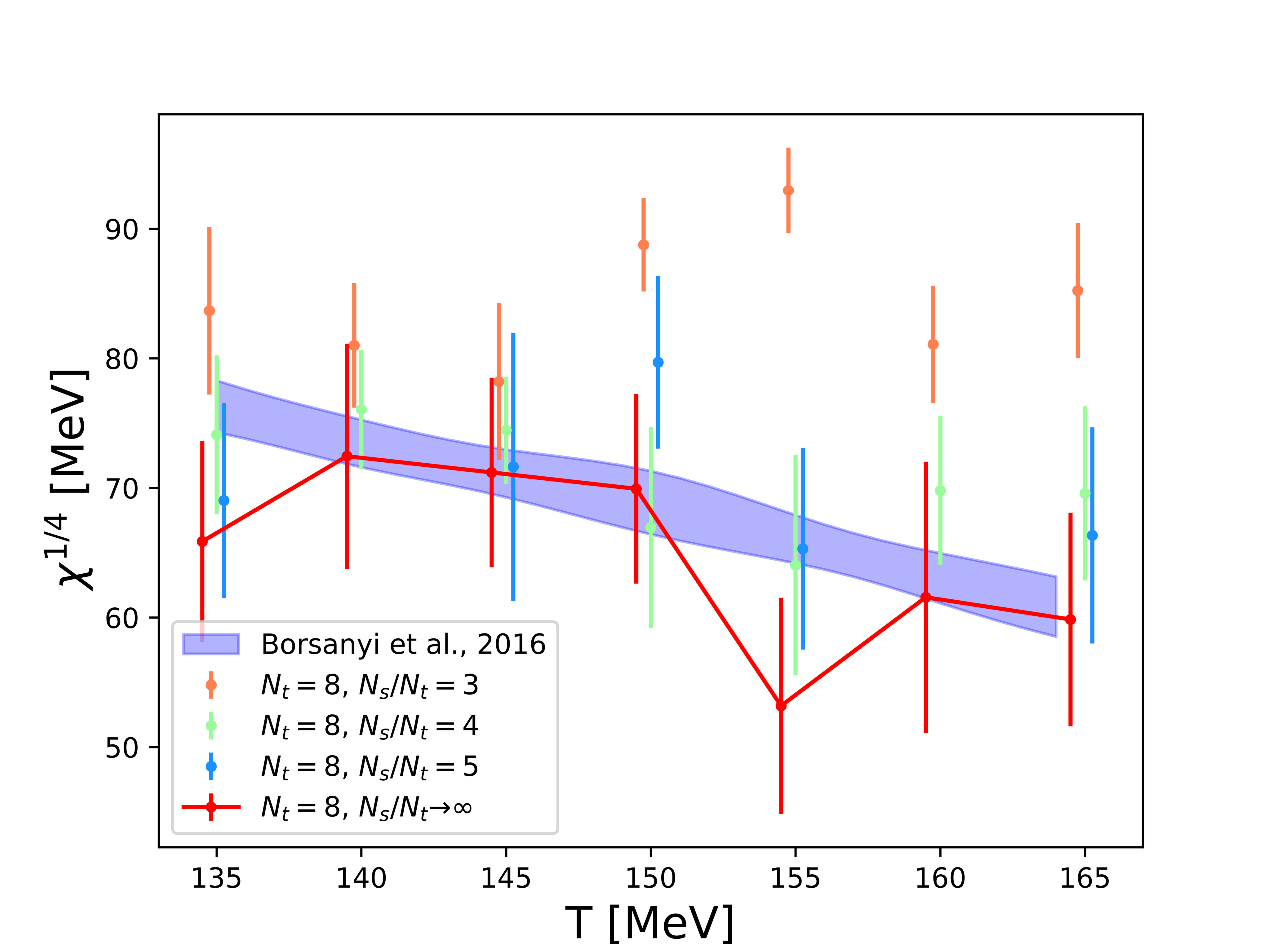}
    \caption{Topological susceptibility $\chi^{1/4}$ as a function of temperature determined using the slab method on all three studied aspect ratios $N_s/N_t=3,4,5$ and its infinite volume extrapolation. Results of \cite{Borsanyi:2016ksw}, based on the staggered fermions, are also presented.}
    \label{fig:topsusslab}
\end{figure}

In Fig.~\ref{fig:slabcorrelator} we present a typical plot for the average $\langle q'^2\rangle$ versus $x$ together with the quadratic fit around $x\sim0.5$. In Fig.~\ref{fig:topsusslab} we present the results for the topological susceptibility $\chi^{1/4}$ determined for all three studied aspect ratios $N_s/N_t=3,4,5$. One can see a nontrivial volume dependence, which we expect to come either from simulations in the Finite Volume or from approximations used in the slab method, Eq.~(\ref{eq:slab}). We performed an extrapolation to infinite volume $N_s/N_t\to\infty$, assuming that finite volume effects scale either as $(N_s/N_t)^2$ (proportional to the surface of the slab) or $(N_s/N_t)^3$ (proportional to the spatial volume) and take the average of these to perform the infinite volume extrapolation as the final result and half of the difference as the systematic uncertainty. In Fig.~\ref{fig:topsusslab} we present also the infinite volume extrapolated result. Also for comparison we show the results for the topological susceptibility, determined in simulations with staggered fermions \cite{Borsanyi:2016ksw}. We observe a perfect agreement between the values for the topological susceptibility determined in both approaches, although our new results, based purely on the simulations with the overlap fermions, have larger uncertainties.

\section{The nature of the phase transition}

In \cite{Fodor:2024gbe} we presented first preliminary results for the chiral observables in the zero topological sector $Q=0$. These results contain information for $N_t=8,10$ and $12$ and several aspect ratios $N_s/N_t=2,3,4$, as well as preliminary data for $N_s/N_t=5$. The main conclusion of \cite{Fodor:2024gbe} is that simulations in the zero topological sector $Q=0$ lead to large finite volume effects, which we expect to be significantly reduced, when one averages the observables for various topological sectors. In order to achieve this goal we generated configurations in several topological sectors $Q=1,2,\ldots$. On these configurations we measured the following chiral observables:
\begin{itemize}
	\item Light quark chiral condensate. We renormalized it using the strange quark condensate:
	\begin{equation}
	\langle \bar{\psi}\psi\rangle_r = m_s \langle \bar{\psi}\psi\rangle_l - 2m_l\langle \bar{\psi}\psi\rangle_s
	\label{eq:cond}
	\end{equation}
	\item Light quark chiral susceptiblity, which is also renormalized by looking at the following combination:
	\begin{equation}
		\chi=m_s\partial \langle \bar{\psi}\psi\rangle_r/\partial m_l
		\label{eq:sus}
	\end{equation}
\end{itemize}

\begin{figure}
    \centering
    \includegraphics[width=0.45\linewidth]{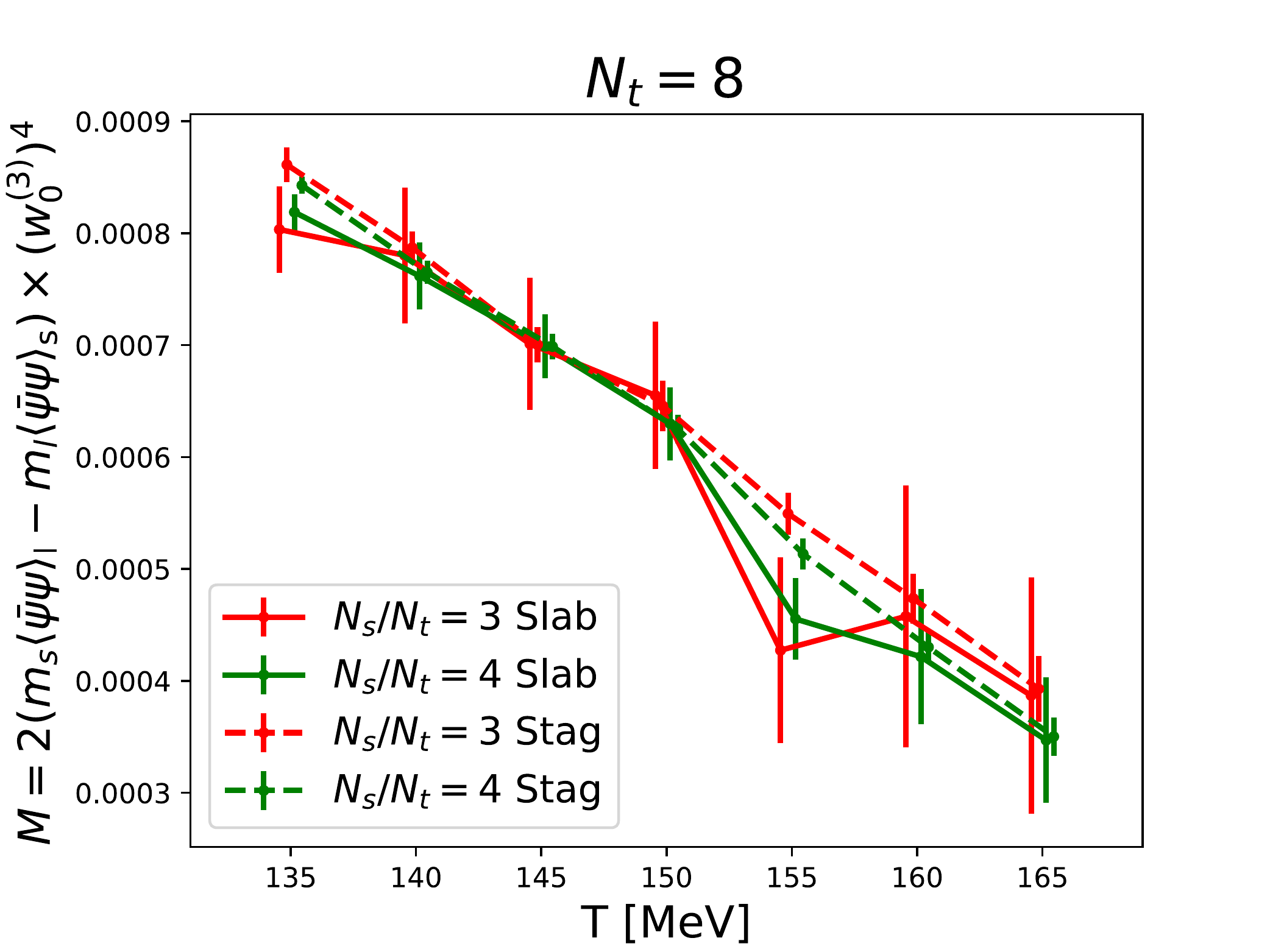}
    \includegraphics[width=0.45\linewidth]{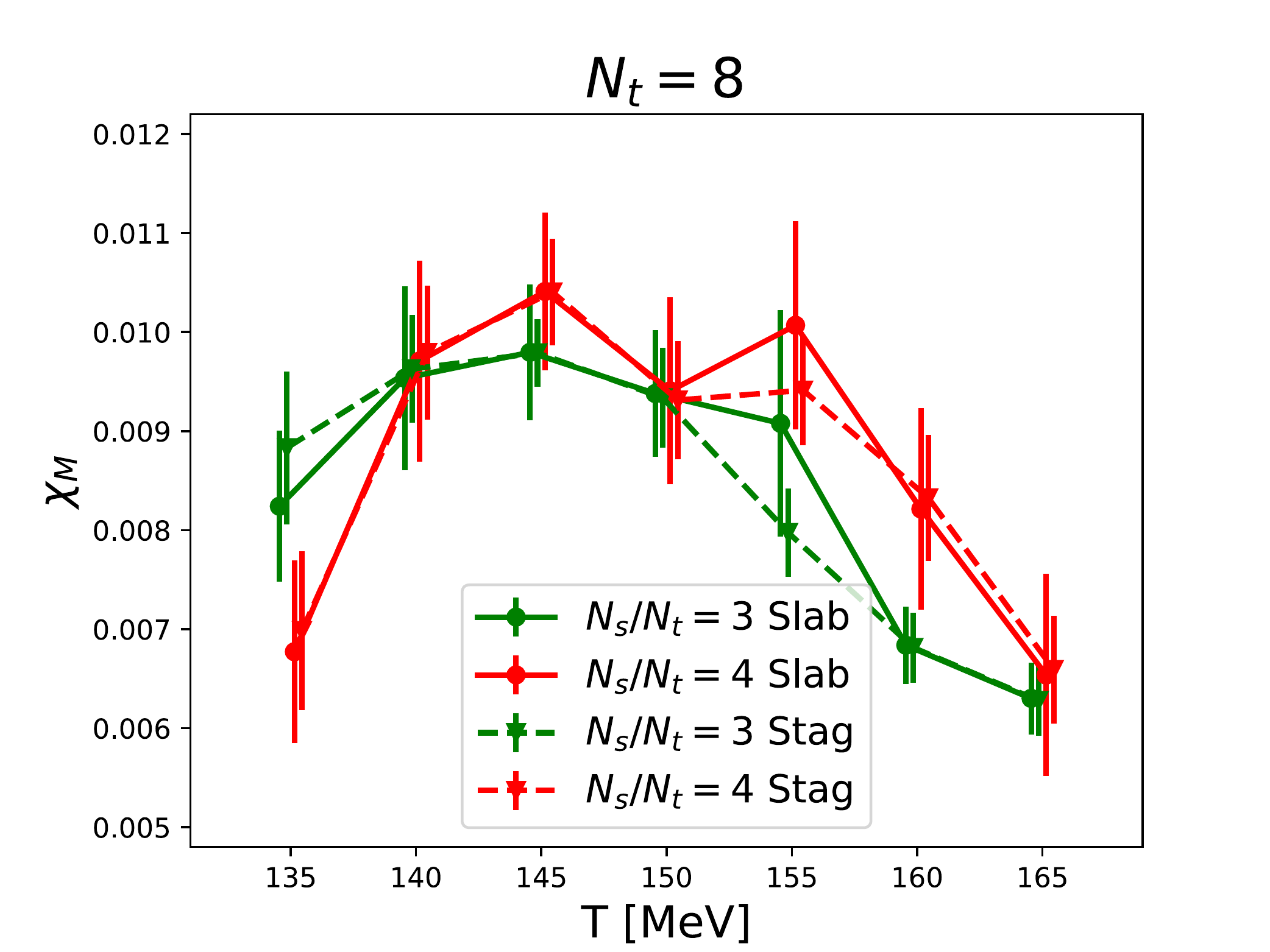}
    \caption{Renormalized chiral condensate (left) and chiral susceptiblity (right) summed over all topological sectors. Red points and lines correspond to the aspect ratio $N_s/N_t=3$, green color represents data for $N_s/N_t=4$. Solid lines correspond to data obtained using the topological susceptibility from overlap simulation and slab method, while for dashed lines we used the topological susceptibility from the staggered based results of \cite{Borsanyi:2016ksw}.}
    \label{fig:summedchobs}
\end{figure}

In order to perform summation over different sectors for these observables, we assumed that the weight of the topological sector $Q$ is given by the Gaussian form $w_Q=Z_Q/Z_0\sim e^{- Q^2/(2\chi V)}$, and we took the value for the topological susceptibility from the slab method, Sec.~\ref{sec:slab}. As a check of our results, we present results summed over all topological sectors using the topological susceptibility from staggered based results \cite{Borsanyi:2016ksw}. Using the weights $w_Q$, we calculate the average of the chiral condensate, Eq. (\ref{eq:cond}) and the chiral susceptibility, Eq. (\ref{eq:sus}) by replacing in these equations the ensemble average $\langle O\rangle$ with the sum $\sum_Q w_Q\langle O\rangle_Q$, where $\langle O\rangle_Q$ represents the average of the observable $O$ in the topological sector $Q$. In Fig.~\ref{fig:summedchobs} we present our results for the chiral observables summed over all topological sectors. We show the results obtained using the topological susceptibility from the overlap configurations using the slab method, also as a check we present the results based on the topological susceptiblity from \cite{Borsanyi:2016ksw}. First, we observe, that results obtained from both values of the topological susceptibility are in agreement with each other, although the results based on the slab method typically have large errorbars. It is naturally a simple consequence of the fact that both results for the topological susceptibility agree with each other and those based on the slab method have larger errorbars, see Fig.~\ref{fig:topsusslab}. Second, we observe the results for the chiral observables for the aspect ratios $N_s/N_t=3$ and $N_s/N_t=4$ are in agreement with each other, in particular the height of the peak for the topological susceptibility is constant for these two aspect ratios, possibly indicating the crossover nature of the phase transition. 

\section{Dirac spectrum}

\begin{figure}
    \centering
    \includegraphics[width=0.45\linewidth]{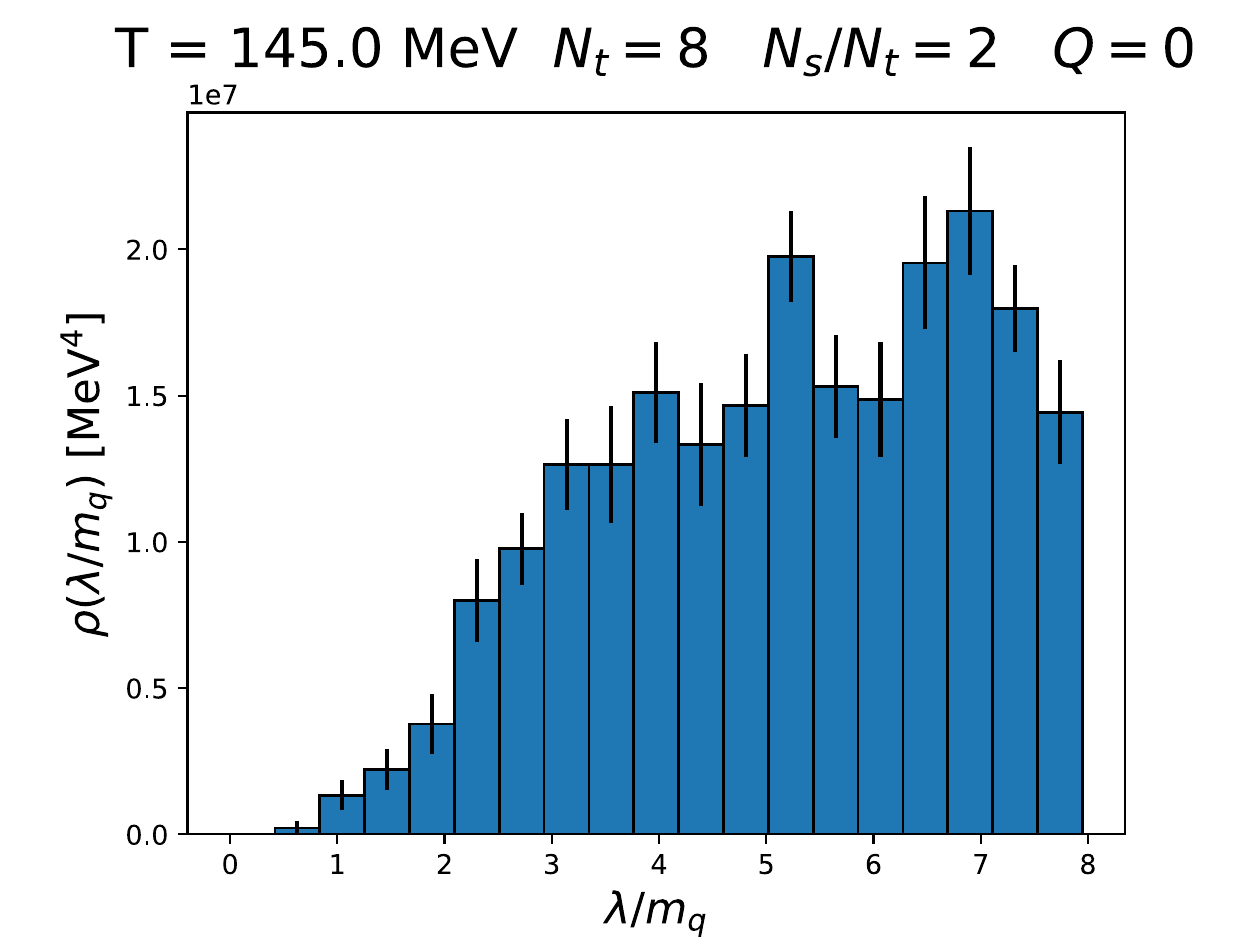}
    \includegraphics[width=0.45\linewidth]{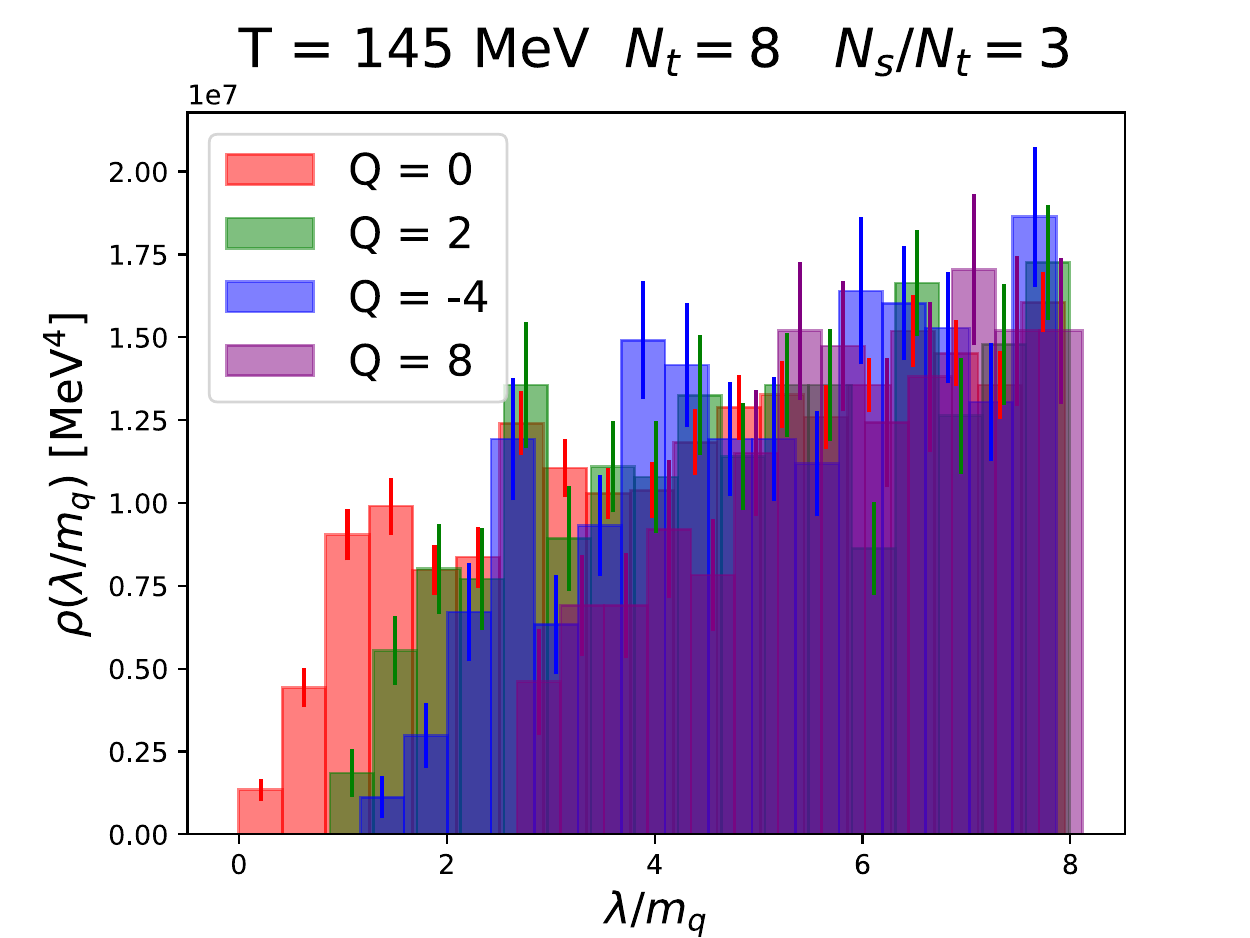}
    \includegraphics[width=0.45\linewidth]{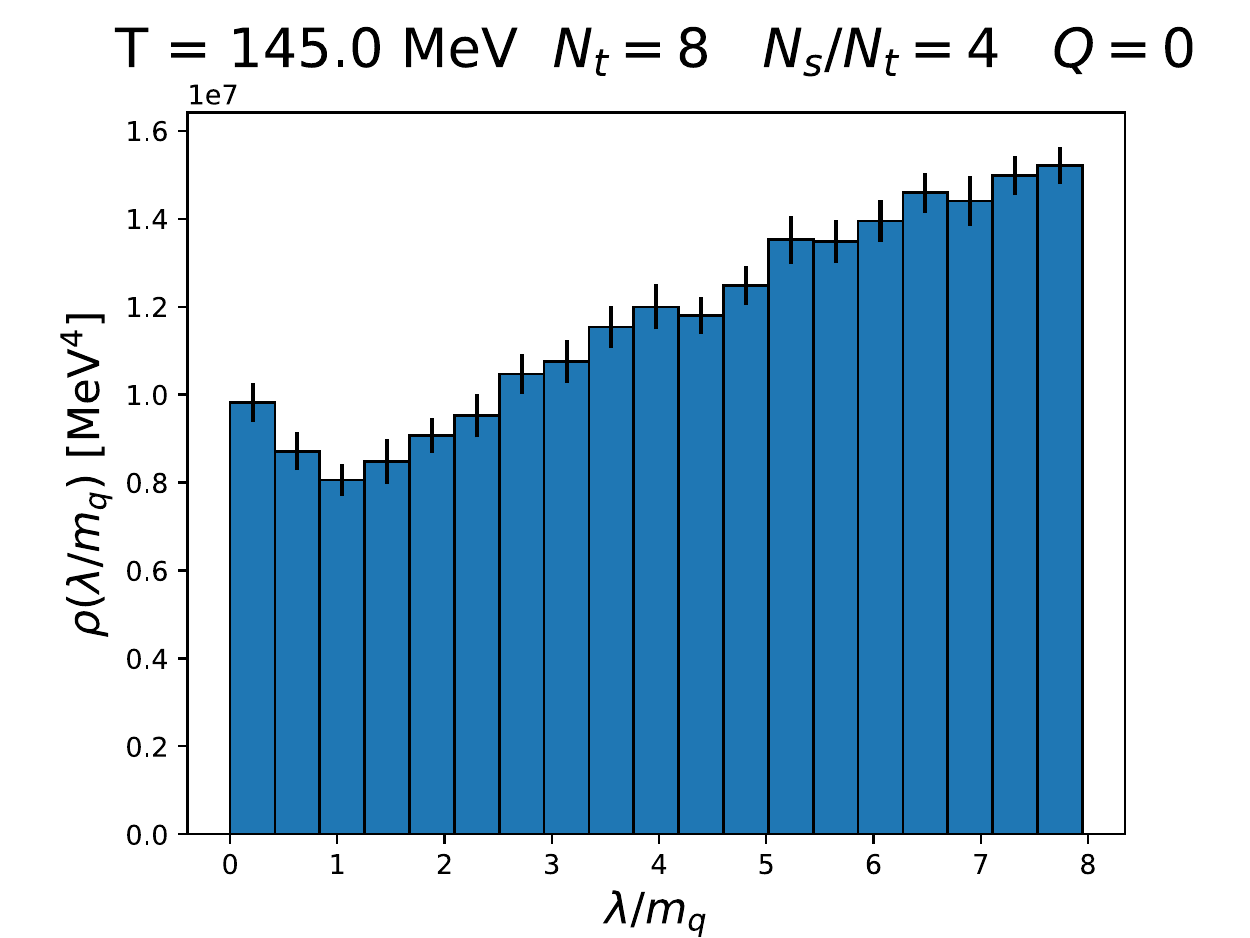}
    \includegraphics[width=0.45\linewidth]{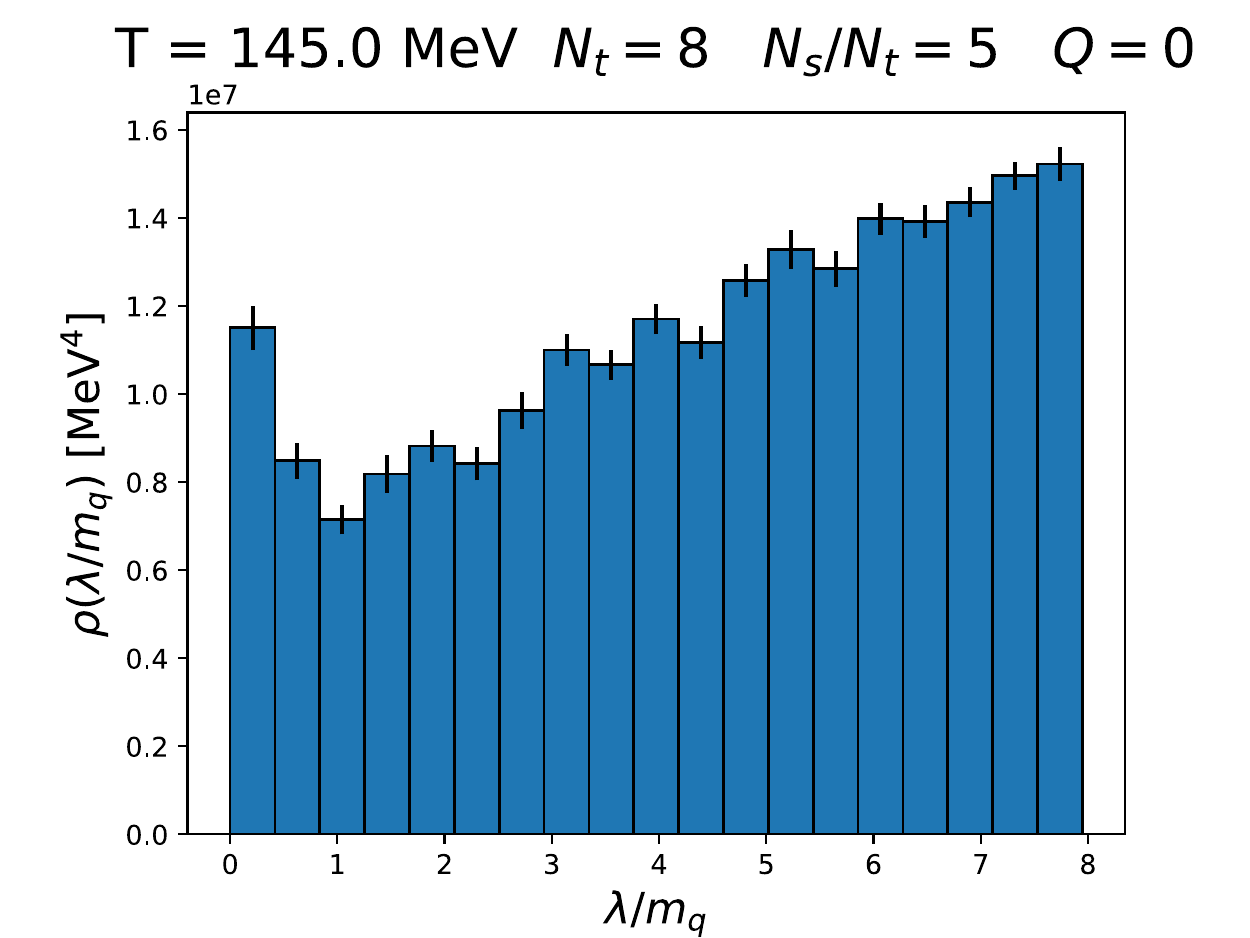}
    \caption{The spectral density of the overlap Dirac operator $\rho(\lambda/m_q)$ for the temperature $T=145$~MeV and four values of the aspect ratio $N_s/N_t=2,3,4,5$ corresponding to different panels. The temporal lattice extent is $N_t=8$. For all aspect ratios the data are for zero topological sector $Q=0$. Additionally for aspect ratio $N_s/N_t=3$ we present data in other topological sectors $=2,-4,8$.}
    \label{fig:eig_145}
\end{figure}

\begin{figure}
    \centering
    \includegraphics[width=0.45\linewidth]{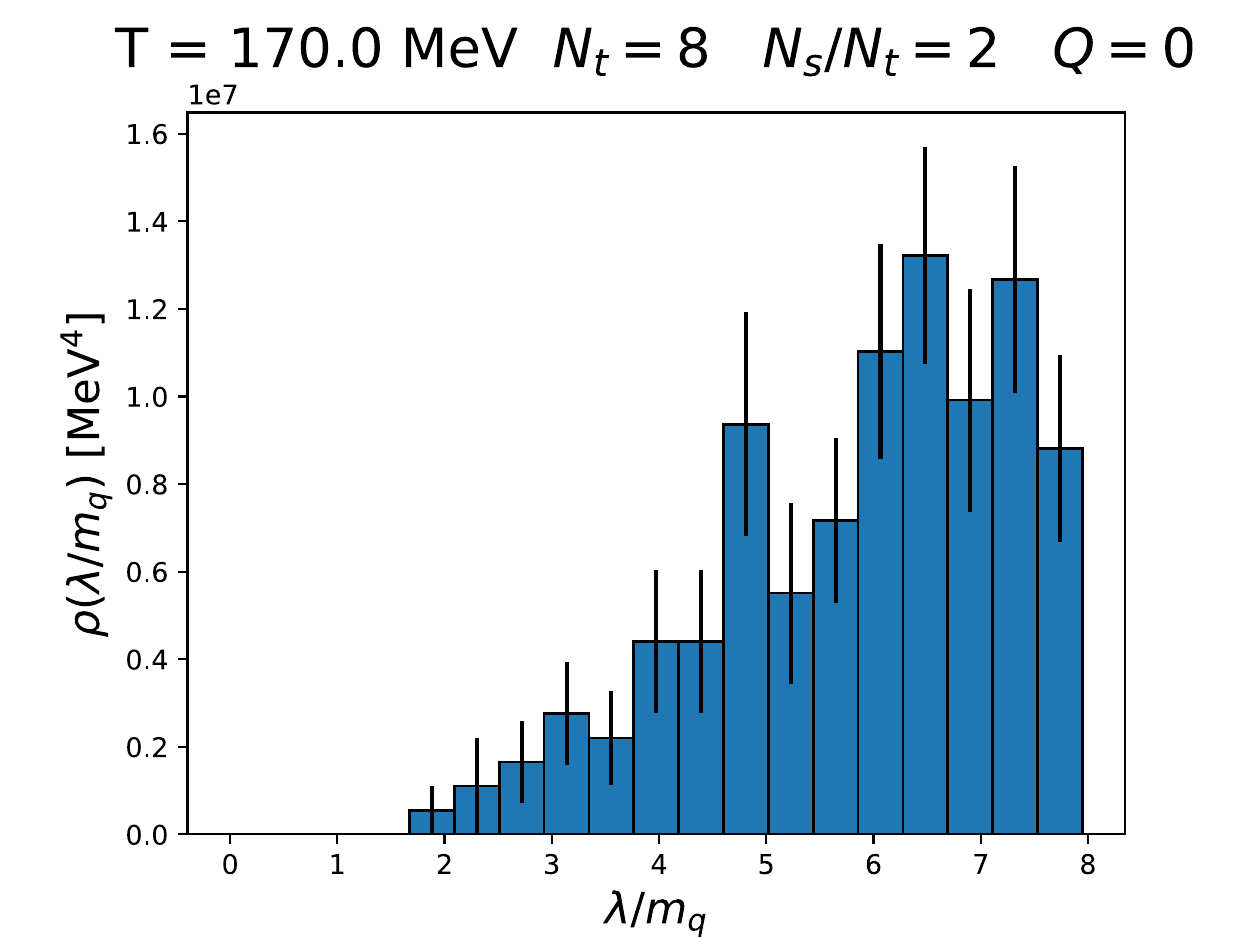}
    \includegraphics[width=0.45\linewidth]{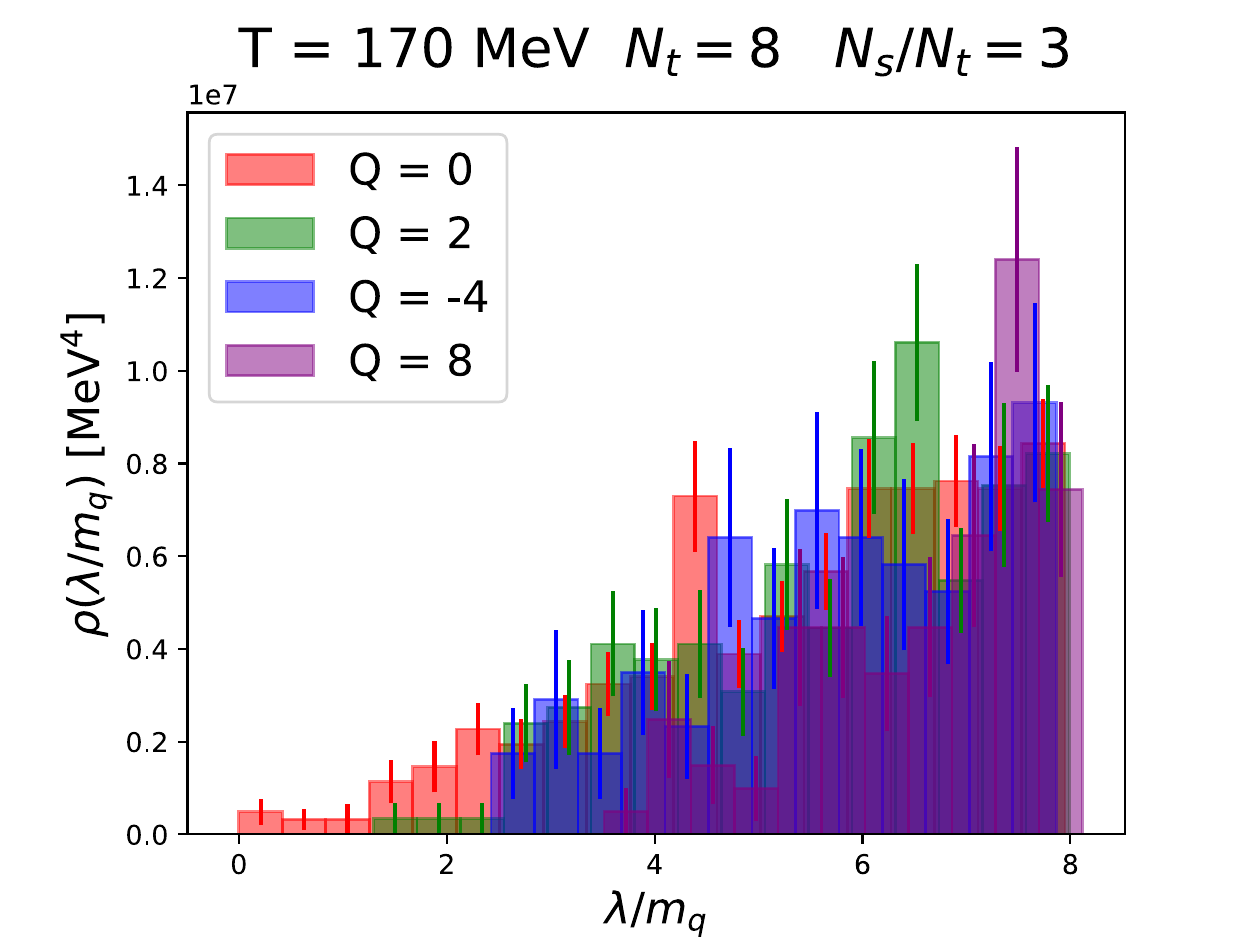}
    \includegraphics[width=0.45\linewidth]{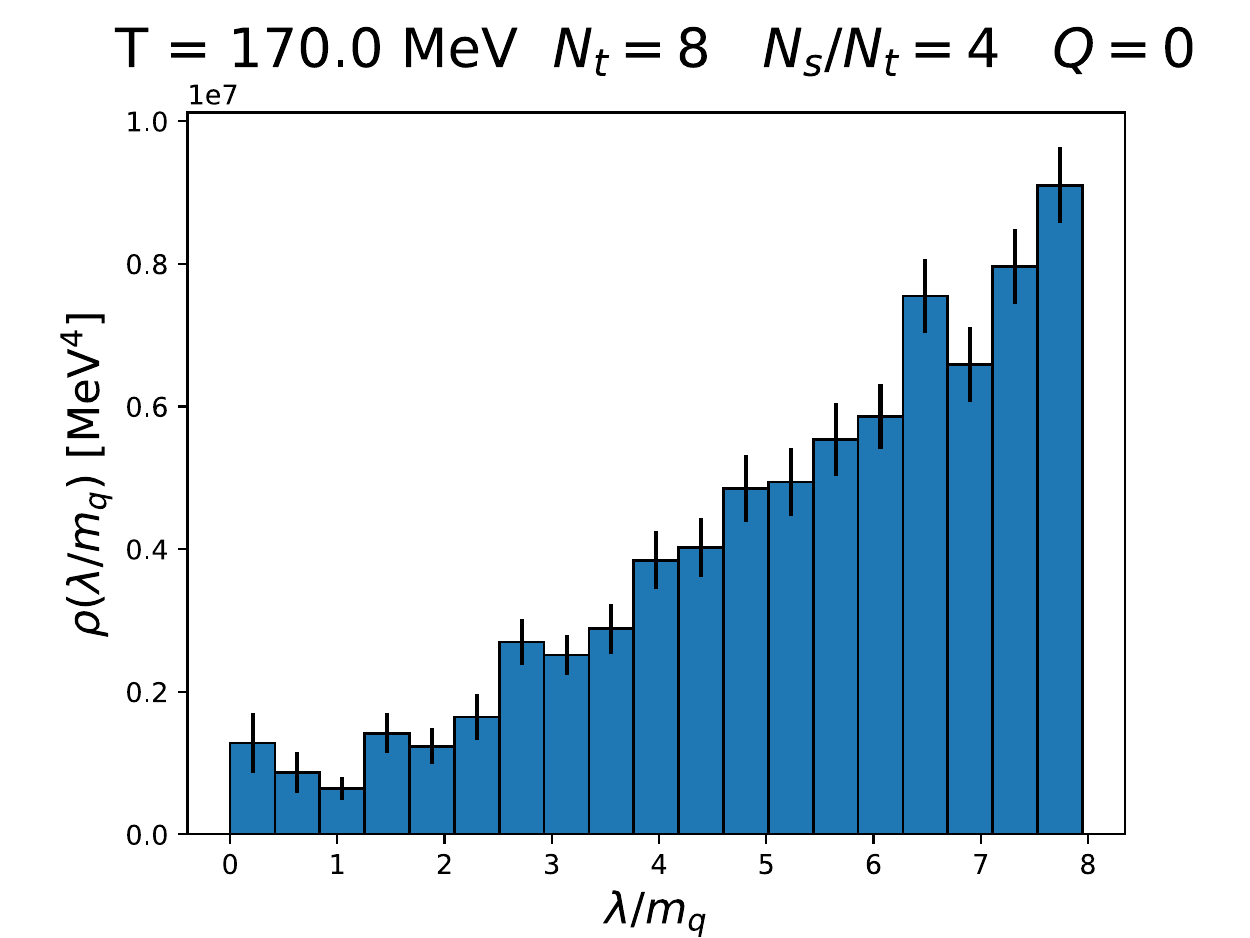}
    \includegraphics[width=0.45\linewidth]{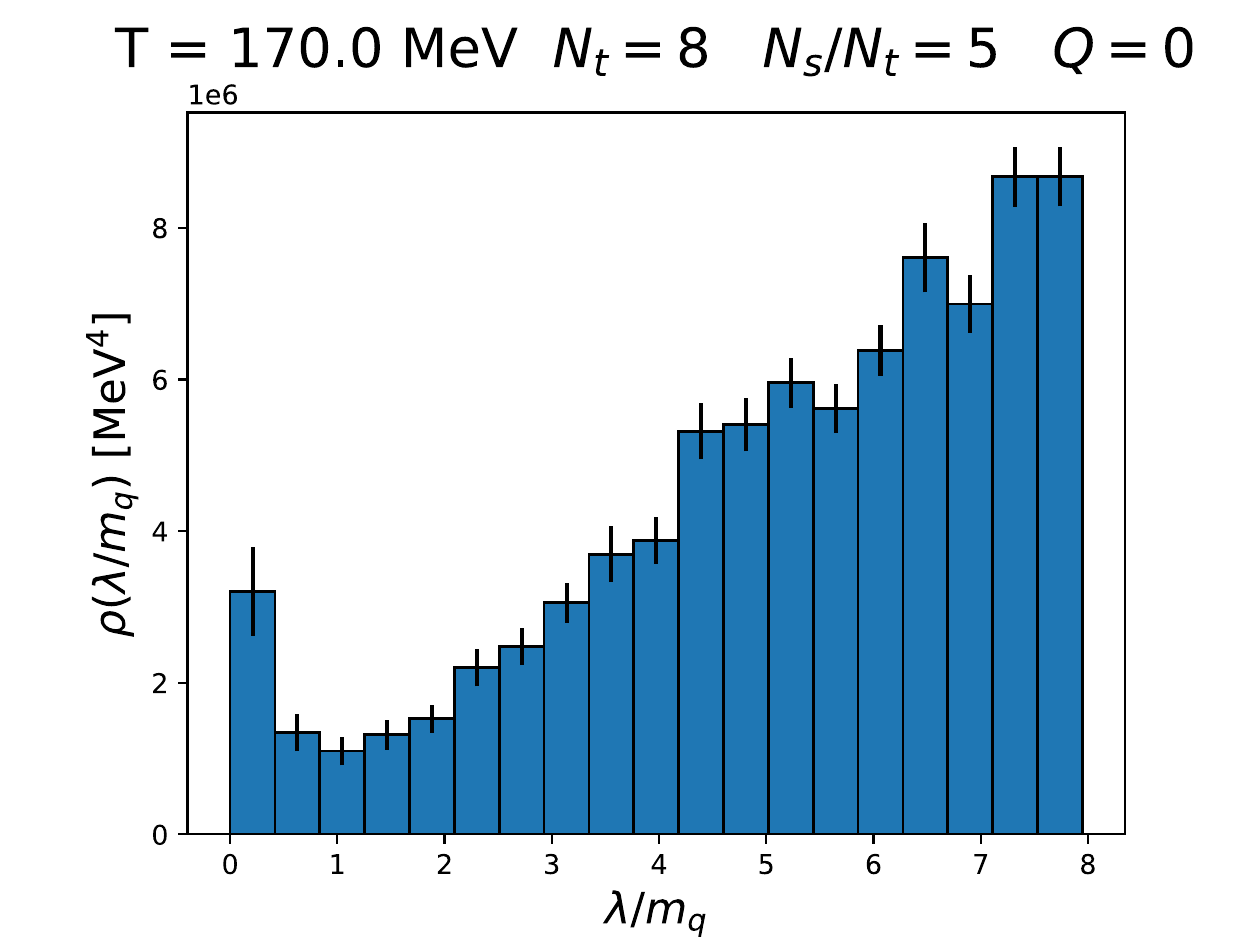}
    \caption{Same as in Fig.~\ref{fig:eig_145}, but for temperature $T=170$~MeV.}
    \label{fig:eig_170}
\end{figure}

One of the interesting observables, which is closely related to QCD symmetries and their behaviour at finite temperature is the spectral density $\rho(\lambda)$ of the Dirac operator. In order to define it, we study the eigenvalues $\lambda_i$ of the overlap Dirac operator $D_{\mathrm{ov}}^{\dag}D_{\mathrm{ov}}$ on the set of generated gauge configurations $U$:

\begin{equation}
    (D_{\mathrm{ov}}^{\dag}D_{\mathrm{ov}})[U] |e_i[U]\rangle = \lambda_i^2[U]|e_i[U]\rangle.
\end{equation}
    
Using the calculated eigenvalues $\lambda_i$, one can construct the spectral density according to:

\begin{equation}
    \rho(\lambda) = \frac{T}{V}\left\langle \sum_i\delta(\lambda-\lambda_i[U])\right\rangle_U,
\end{equation}
    
where $\langle\rangle_U$ represents averaging over gauge field configurations. The chiral condensate $\langle\bar{\psi}\psi\rangle$, which is the order parameter of the chiral phase transition, can be determined from the $\rho(\lambda)$ using the Banks-Casher relation \cite{Banks:1979yr}:

\begin{equation}
    \langle\bar{\psi}\psi\rangle = \int \frac{m}{\lambda^2+m^2}\rho(\lambda)d\lambda \xrightarrow[m\to0]{}\rho(\lambda).
\end{equation}

At the same time, the order parameter for the axial $U(1)_A$ symmetry is the axial susceptibility $\chi_A$, which is given by the  difference between susceptibilities defined from the correlators in the $\pi$ and $\delta$ channels and is also related to the spectral density $\rho(\lambda)$ \cite{Chandrasekharan:1998yx}:

\begin{equation}
    \chi_A = \chi_{\pi}-\chi_{\delta} = \int \frac{m^2}{(\lambda^2+m^2)^2}\rho(\lambda)d\lambda.
\end{equation}
     
These relations stress the connection between the Dirac spectrum and QCD symmetries.

It is known that the eigenvalues $\lambda$ are multiplicatively renormalized with the same factor as the quark mass $m_q$, thus the ratio $\lambda/m_q$ is RG invariant \cite{Kaczmarek:2021ser,Giusti:2008vb}.  For this reason we study the spectral density as a function of $\lambda/m_q$. In Fig.~\ref{fig:eig_145} we present the spectral density $\rho(\lambda/m_q)$ as a function of $\lambda/m_q$ for the temperature $T=145$~MeV, several aspect ratios $N_s/N_t=2,3,4,5$. We mainly present data in the zero topological sector $Q=0$. Additionally, for the aspect ratio $N_s/N_t=3$ we show the spectrum in other topological sectors $Q=2,-4,8$. Note that the temperature $T=145$~MeV is slightly below the pseudocritical temperature. For small aspect ratios $N_s/N_t=2,3$, the zero virtuality limit of the spectral density $\rho(\lambda/m_q\to0)$ is consistent with zero\footnote{Note, that simulations in the non-zero sectors show even larger suppression of the near zero modes.}. This is clearly a finite volume effect, as for larger aspect ratios $N_s/N_t=4,5$ one clearly sees  a non-zero intercept $\rho(\lambda/m_q\to0)$. Moreover, it can be easily seen that the spectral density has a peak at $\lambda\to0$. The effect becomes even more pronounced for the higher temperature $T=170$~MeV, just above the chiral transition, where the intercept $\rho(\lambda/m_q\to0)$ is zero or very small for all aspect ratios, except for the largest $N_s/N_t=5$, on which we clearly see a peak-like behaviour in the limit $\lambda\to0$. 

The existence and behaviour of this peak-like structure in the spectral function has been the subject of multiple studies. Several lattice calculations carried out with various fermion discretizations \cite{Kaczmarek:2021ser,Alexandru:2015fxa,Ding:2020xlj,Alexandru:2024tel,Alexandru:2023xho,Aoki:2020noz} tend to converge at least qualitatively to the conclusion that this peak is not a lattice artifact. In~\cite{Alexandru:2019gdm} it was also argued that this behaviour might lead to a new phase in QCD.
In~\cite{Kovacs:2023vzi} an instanton-based random matrix model was proposed, which explains the spectral density and the peak $\rho(\lambda\to 0)$ via a gas of free instantons and anti-instantons. Our results are in agreement with this mechanism.

\section{Summary}

In this Proceeding we presented results of our study of the thermal QCD transition using the overlap fermion discretization with temporal lattice extent $N_t=8$. We generated configurations in fixed topological sectors for several values of the topological charge $Q$, and aspect ratios $N_s/N_t=3,4,5$ and measured chiral observables. In these fixed topological sector simulations we measured the topological susceptibility using the slab method. With the help of the determined values of the topological susceptibility we averaged the chiral observables over all topological sectors and determined their full temperature dependence. We would like to stress that these results are obtained purely in the simulations with the dynamical overlap fermions. The volume dependence of the chiral susceptibility is consistent with the crossover nature of the thermal QCD transition. Additionally we measured the spectrum of the overlap Dirac operator, which turns out to be in the agreement with the chiral observables. Remarkably, we confirm the presence of a pronounced peak in the Dirac spectrum at $\rho(\lambda\to 0)$, even with dynamical overlap quarks. However, our results also show that even close to the critical temperature, large aspect ratios ($N_s/N_t=4$ and  $N_s/N_t=5$) are needed to observe the peak. If the instanton-based explanation of the peak is correct, at higher temperatures much larger volumes would be needed to capture the peak, as the instanton density falls sharply with the temperature.

\section{Acknowledgements}

The authors gratefully acknowledge the Gauss Centre for Supercomputing e.V. (www.gauss-centre.eu) and VSR Commission for providing computing time on the GCS Supercomputer  JUWELS \cite{JUWELS} and on the supercomputer JURECA \cite{JURECA} at Jülich Supercomputing Centre (JSC).

\bibliographystyle{unsrt}
\bibliography{bib}

\end{document}